# Racket Kinematics and Ball Performance During the Forehand Reverse Flick: A Pilot Comparison With the Backhand Flick in Table Tennis

Pengfei Jin[1], Jie Ren[2], Chen Yang[3], Qingtao Kong[1], Qingshan Zhang[4], Nan Gu[2], Bin Chen[2], Qin Zhang[2], Zhe Feng[2*]

[1]Department of Physical Education and Sport, Shanghai Ocean University, Shanghai, China, [2]China Table Tennis College, Shanghai University of Sport, Shanghai, China, [3]SWJTL-LEEDS Joint School, Southwest Jiaotong University, Chengdu, China, [4]School of Athletic Performance, Shanghai University of Sport, Shanghai, China

Correspondence: Zhe Feng (fengzhe202203@163.com)

**Abstract:** This study aimed to evaluate the forehand reverse flick (FRF), an offensive return technique for short balls to the forehand, and compare it with the backhand flick (BF) after four elite table tennis players completed seven consecutive days of FRF-specific training. Infrared motion capture and ultra-high-speed cameras were used to collect data on racket kinematics, movement duration, and ball performance. The success rate of the FRF increased steadily, reaching 86%. Racket trajectories of the two techniques were highly similar along the X (r = 1) and Y (r = 0.99) axes but differed along the Z (r = -0.04) axis. Racket and ball velocities were comparable between techniques, whereas the FRF showed lower resultant acceleration (approximately 265.57 m/s²) and required about 0.03 s more for movement duration. Ball velocity was comparable between techniques, for the ball spin, the FRF generated lower spin (approximately 76.61 r/s) about 64% of the BF value (approximately 120.13 r/s). The highest participant mean spin rate reached 93 r/s, about 77% of the BF mean. Overall, the FRF was found to have favorable learnability and training value, with potential for further optimization and competitive application.

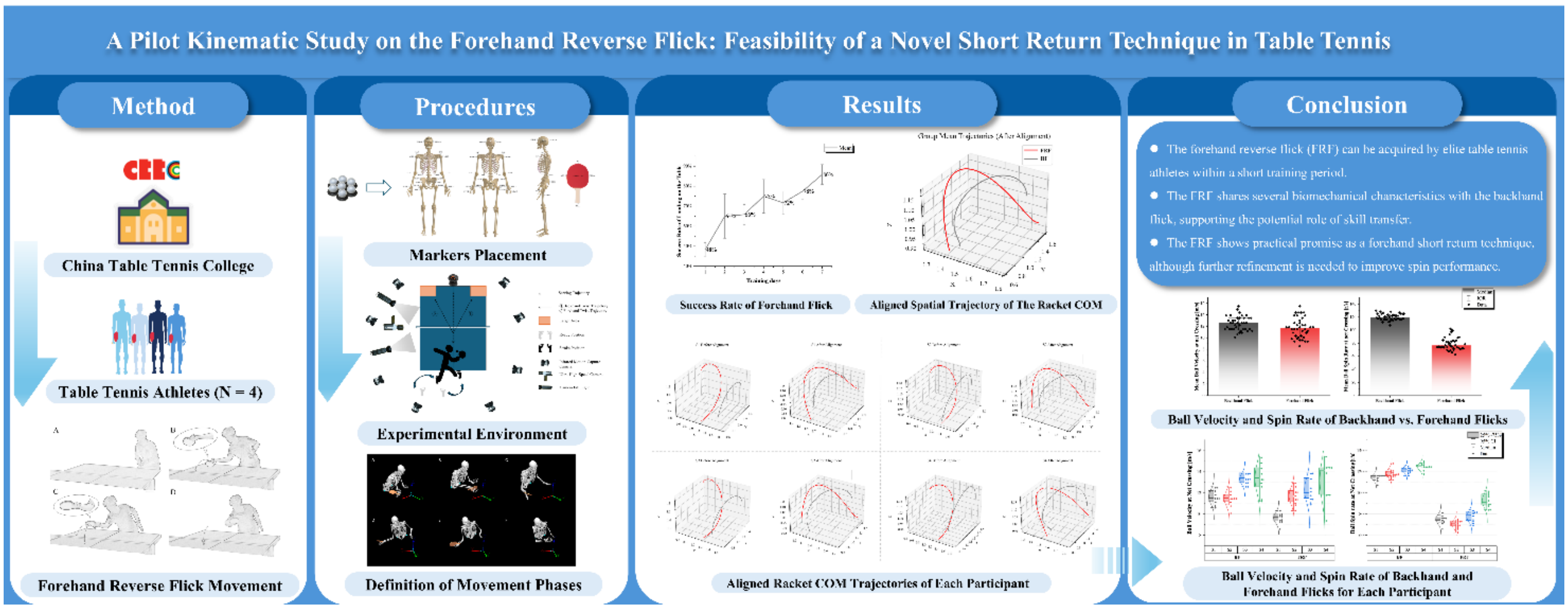

**Graphical Abstract**

## Introduction

Since 2014, the International Table Tennis Federation (ITTF) has replaced celluloid balls with 40+ plastic balls. Compared with celluloid balls, plastic balls generally exhibit lower flight velocity and spin rate, along with different rebound and frictional characteristics upon contact with the table[1,2]. These changes have subtly reshaped the technical–tactical ecology of high-level matches, amplifying the importance of the first three strokes. In particular, the quality of the return has become critical for either scoring directly (won directly with a return) or initiating subsequent topspin rallies. Previous studies have reported that the introduction of plastic balls reduced the effectiveness of the server as a direct scoring technique, while points won by return winners and counterattack combinations accounted for approximately 30%. Concurrently, the frequency of the backhand flick—a backhand attacking technique primarily employed in service return—has increased[3,4].

During service return, the backhand flick (BF) generates active topspin to secure sufficient clearance over the net and greater ball depth, thereby increasing the difficulty of the return of the opponent and enhancing the threat of the receive. A high-quality backhand flick relies on coordinated force production between the trunk and upper limb[5]. Specifically, during the backswing, the elbow is raised, the forearm internally rotates, and the racket is drawn back in front of the abdomen. In the forward swing, based on core stability, the lower limb of the athlete pushes off the ground and rotates the trunk, while the forearm rapidly externally rotates and accelerates forward–upward to strike the ball, brushing the mid-posterior lateral surface[6,7]. Table tennis studies have shown that the racket head rotation angle in the horizontal plane is closely associated with the magnitude of topspin generated[8]. During the forward swing, with the elbow joint serving as the axis of rotation, the combined action of the shoulder, elbow, and wrist guides the racket head along a horizontal rotational path, allowing the line of action of the racket force to deviate from the center of mass of the ball. This deviation produces a pronounced tangential force component, thereby enhancing the ability to generate topspin to the ball[9].

With respect to technical–tactical applications in competition, Yang et al. analyzed 258 matches of elite athletes using a shot effectiveness model and found that, for both male and female players, the second and fourth strokes (approximately 35%) were more decisive than the first and third strokes (approximately 25%)[10]. This pattern arises because the backhand flick has been more effectively applied to short serves directed to the backhand or middle areas when using the new plastic ball[11]. In particular, the younger generation of male players tend to employ the backhand flick more frequently, with representative athletes including Fan Zhendong (China), Harimoto Tomokazu (Japan), and Hugo Calderano (Brazil). However, this trend also exposes a technical gap: when the serve is placed short to the forehand, players face the challenge of maintaining accuracy while consistently producing threatening topspin returns within a limited temporal window and restricted spatial position.

At present, forehand short offensive return techniques can be broadly classified into two categories. The first involves applying the backhand flick from the forehand side, which has the advantage of producing threatening backhand flick returns but substantially increases the difficulty of defensive recovery to the ready position. The second is the use of the forehand flip to complete the return[6], which is characterized by a compact stroke movement and fast horizontal velocity of the ball, but it lacks the ability to generate strong topspin, thereby failing to balance power and stability. Regarding technical innovation in forehand short returns, Xiao et al. analyzed the innovative forehand flip–loop combination technique inside the table with penhold athletes in their study. Their results indicated that this technique produced higher spin rates than the forehand flip[12]. However, the serve placement in that study was set to the backhand short zone, and the stroke was executed with lateral wrist force from a pivoted body position, rather than simulating forehand short placements as in actual matches. Concerning the forehand flick, the penhold forehand flick of Xu Xin relies on wrist adduction and external rotation during the backswing, followed by rapid abduction and internal rotation to execute a lateral flick. Although this technique can serve to exploit the unpreparedness of the opponent, it is primarily dependent on wrist action. Due to the anatomical limitations of the radiocarpal joint, the forward horizontal velocity imparted to the ball is insufficient. While this stroke is capable of generating sidespin, its overall threat and stability are limited, which explains why it has not been widely adopted. The definitions of the short return techniques used in this study are summarized in Table 1 to ensure terminological clarity.

Table 1. Definitions of short return techniques

| Technique | Definition |
|---|---|
| Backhand Flick | A backhand short return technique executed with brushing contact, primarily aimed at generating strong topspin. |
| Backhand Flip | A backhand short return technique executed mainly through impact contact, producing a fast but low-spin return. |
| Forehand Flip | A forehand short return technique executed mainly through impact contact, producing a fast but low-spin return. |
| Forehand Reverse Flick | A novel forehand short return technique executed with the backhand side, in which brushing contact is used to generate topspin from the forehand side. |

Based on the limitations of existing forehand short-play techniques, the present study proposes a forehand short return technique, termed the forehand reverse flick (FRF). This technique involves forearm external rotation to turn the racket into the backhand side at the forehand short-ball position, with the racket swing trajectory mimicking that of the backhand flick. The forehand flick resembles the forehand stroke using a semi-western grip in tennis, which is known to produce substantial topspin during groundstrokes[13]. From a kinematic perspective, shakehand athletes can achieve a horizontal racket angle similar to that of the backhand flick at the end of the backswing through forearm external rotation and wrist abduction. During the forward swing, shoulder adduction, forearm pronation, and wrist flexion combine to generate high racket velocity, as illustrated in Figure 1. Accordingly, the present study examined elite shakehand athletes through training and kinematic testing to compare the similarities and differences between the FRF and BF. The study proposed three hypotheses: 1) elite athletes would acquire the FRF technique efficiently through short-term training; 2) the FRF and BF would exhibit comparable racket kinematic characteristics; and 3) the FRF would produce lower ball velocity and spin rate than the BF. The aim was to evaluate the feasibility and practical value of the FRF in forehand short returns, thereby addressing the limitations of existing forehand short-return techniques and contributing to a breakthrough in the evolution toward an era dominated by topspin in table tennis.

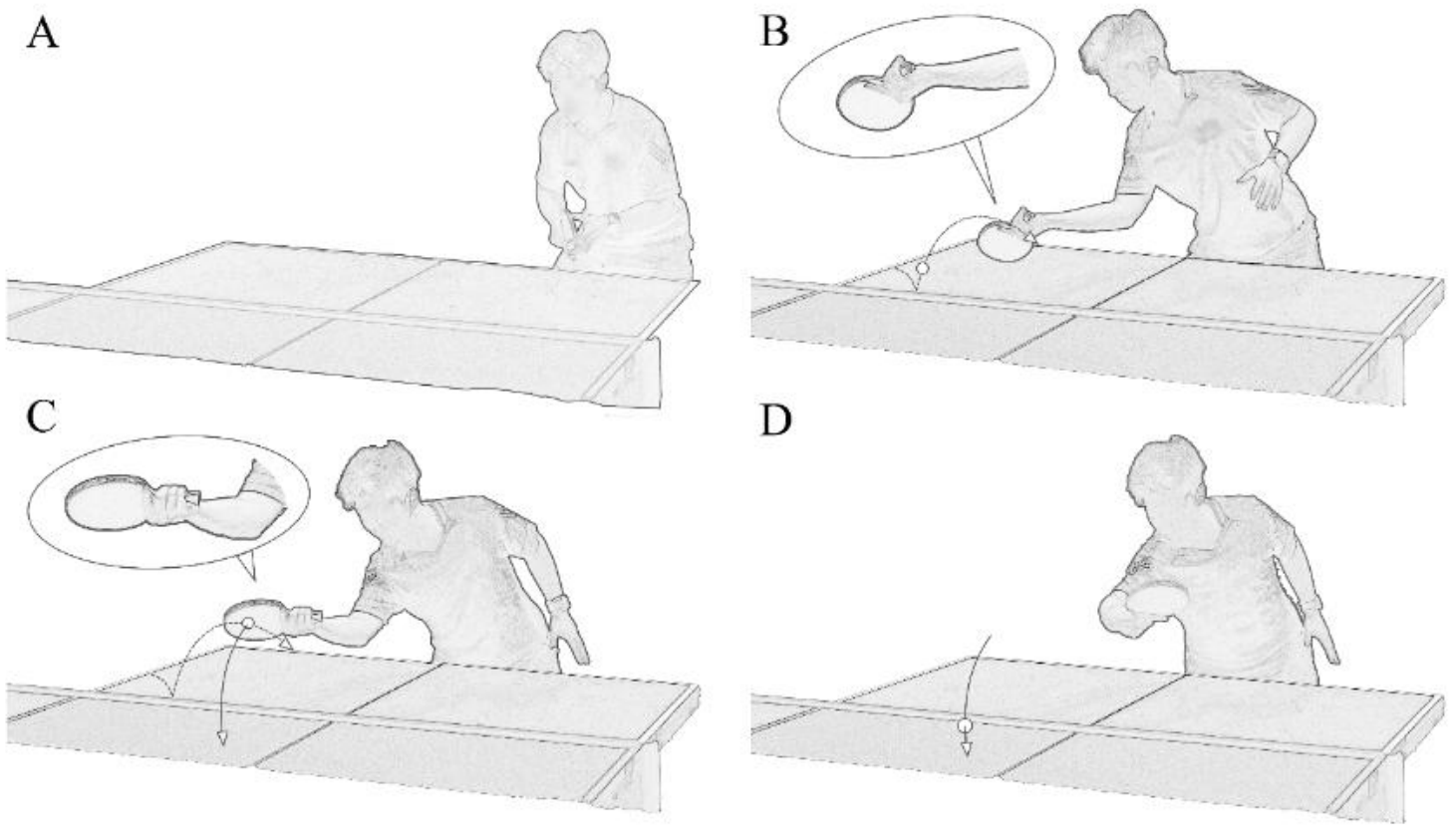

Figure 1 Demonstration of the forehand reverse flick technique

## Materials and methods

### Subjects

Four elite male athletes participated the present study (age: 21.25 ± 0.5 years, height: 1.80 ± 0.04 m, weight: 67.84 ± 6.73 kg). Participants were recruited from the highest-level training group (Group A) of the Table Tennis College, Shanghai University of Sport. The sample included two national first-class athletes and two national master-level athletes. Inclusion criteria were as follows: (1) right-handed shakehand grip; (2) absence of upper- or lower-limb disorders; and (3) no injury within six months before testing. Exclusion criteria were as follows: (1) consumption of alcohol- or caffeine-containing beverages within 24 h prior to testing; and (2) presence of joint pain or movement disorders on the day of testing. All participants provided written informed consent before participation. The study protocol was approved by the Ethics Committee of Shanghai University of Sport (102772025RT115) and conducted in accordance with the ethical principles outlined in the Declaration of Helsinki.

### Design and Procedures

A pilot, nonrandomized, within-subject repeated-measures design was employed. The basic framework and training program of the FRF were established after consultation with several national team coaches. Selected athletes underwent one week of training, with 30 minutes per day. Because the FRF can benefit from skill transfer of the backhand flick and because elite athletes generally demonstrate strong learning capacity and technical comprehension, the training focused on 30 minutes of FRF practice per day, either in isolation or in combination with other techniques. After the warm-up, participants trained in pairs, and the seven-day training contents are presented in Supplementary Table 1.

Before the first training session, participants were given standardized instructions: the FRF can be regarded as a technical transfer from the backhand flick, differing from the forehand flip in that it emphasizes spin generation through brushing contact. The key aspects of spin production were highlighted as follows: 1) Backswing distance and racket-head trajectory: During the backswing, shoulder abduction, forearm external rotation, and wrist abduction should position the racket head pointing backward at the end of the preparation phase, thereby increasing the amplitude and spatial allowance of the swing. The racket trajectory should form a semi-elliptical accelerating curve. 2) Racket velocity: In the forward swing, rapid execution of shoulder adduction, forearm pronation, and wrist flexion is required to accelerate the racket. It is important to control wrist motion during acceleration to avoid excessive movement. 3) Racket angle at ball–racket contact: At the moment of impact, the racket angle should approximate that of the backhand flick. After contact, the arm should follow through, and the body should quickly recover to the ready position.

After each daily FRF training session, the success rate of the FRF was assessed using a ball-feeding robot (HALO PLUS V1.2, Pongbot, China). The robot was set to the following parameters: lateral displacement 5, speed 1.5, spin 7.0, frequency 30%, depth 20, backspin, micro-adjustment 0, and number of balls 33. For each test, the first three balls were used for familiarization, and the success rate of the FRF was recorded for each participant after completing seven days of training.

**Experimental Procedures**

Eight infrared motion capture cameras (Qualisys MRI, Gothenburg, Sweden) were used to collect kinematic data of joint marker points at a sampling frequency of 200 Hz. Prior to data collection, spatial coordinate system positioning and spatial calibration of the experimental environment were conducted using an L-shaped calibration frame and a T-shaped wand, respectively. The calibration protocol required the residual error of each camera to be no greater than 0.8 mm, and the overall standard deviation across all eight cameras to be below 0.8 mm. Cameras that exceeded the calibration threshold were masked and recalibrated until the required standard was achieved.

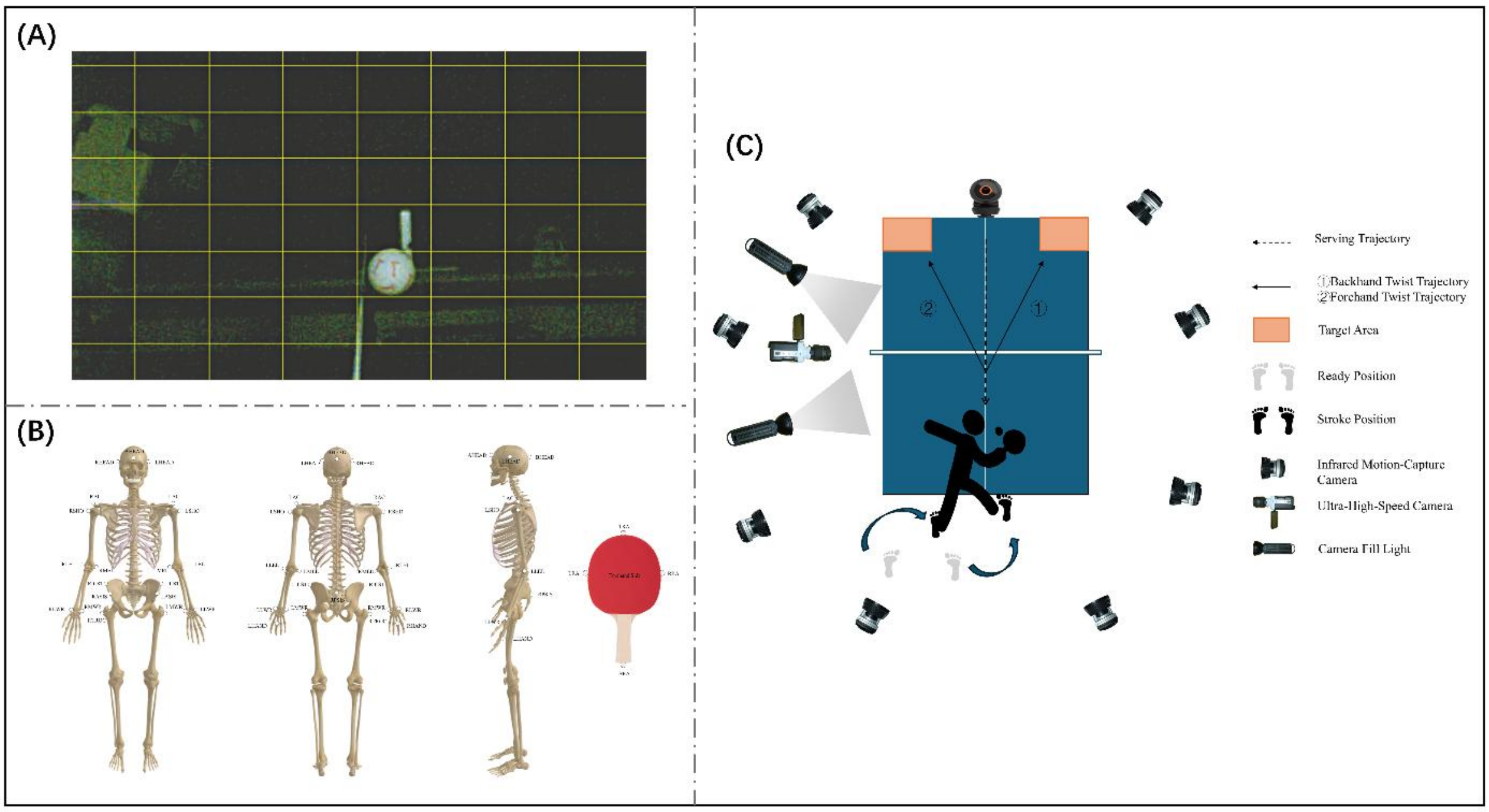

Figure 2 Image acquisition, marker placement, and experimental setup for the kinematic and ball-return quality test

Note: (A) Image acquisition using the high-speed motion capture camera. (B) Markers Placement. (C) Schematic diagram of the kinematic experiment and ball-return quality test setup.

A Phantom Miro LC111 8002 ultra-high-speed camera (Vision Research Inc., Wayne, New Jersey, USA) was positioned approximately 30 cm from one side of the table. Two supplementary lights were placed on both sides of the camera to increase indoor illumination. The camera view was adjusted so that the center of the frame aligned with the table net line. A ball marked with the letter "T" was positioned above the net height for focusing. Camera parameters were set to a sampling frequency of 10,000 fps, a resolution of 512 × 320, an exposure time of 99.00 μs, and a recording range of [−17338, 0]. The position and focal length of the lens were adjusted until the "T" mark on the ball could be clearly observed after focusing, as shown in Figure 2A. Because the extremely high sampling frequency limited the number of frames that could be stored, the ball quality test was performed after the completion of the kinematic measurements. The ball's crossing spin rate was calculated based on the number of frames required for the "T" mark to complete one full rotation. The crossing ball velocity was calculated from the number of frames required for the ball to travel a distance equivalent to one ball diameter (d = 40.3 mm).

All participants wore standardized black tight-fitting clothing. After the technical training, a total of 25 reflective infrared markers (diameter: 44 mm) were attached to specific anatomical landmarks of the athletes: iliac crests (bilateral), anterior superior iliac spines (bilateral), posterior superior iliac spines (on the line connecting both iliac crests), greater trochanters of the femur (bilateral), acromions (bilateral), greater tubercles of the humerus (bilateral), medial and lateral epicondyles of the humerus (bilateral), radial styloids (bilateral), ulnar styloids (bilateral), and third metacarpal heads (bilateral). In addition, four markers were attached to the racket, as illustrated in Figure 2B.

Before testing, each participant performed a 10-minute standardized warm-up guided by the same coach. The warm-up consisted of 3 minutes of upper- and lower-limb stretching, 3 minutes of forehand and backhand loop practice, and 4 minutes of FRF and BF practice. The experimental task involved executing FRF and BF against backspin balls. The ball-feeding machine delivered double-bounce backspin balls consistently toward the centerline of the table, with the following settings: lateral displacement 3, speed 1.5, spin 6, frequency 10%, depth 20, underspin, micro-adjustment 0. The kinematic testing protocol required participants to:1) adopt their personal ready position as in competition; 2) perform the designated FRF or backhand flick and return rapidly to the ready position after ball contact; 3) attempt to return the ball with maximum effort to a designated cross-court target area (30 × 30 cm) with the ball successfully landing on the table. A stroke was considered valid only if all requirements were met. Data were collected in the order of BF followed by FRF. After each set of one technique, participants rested for 2 minutes. Each technique was performed 20 times, resulting in 40 trials in total (2 techniques × 20 trials = 40). as illustrated in Figure 2C. For data analysis, five valid trials were selected for each participant, and the kinematic variables were averaged.

Based on previous studies, the motion of the racket in the horizontal plane was used as the primary observation reference, and the stroke was divided into three key events and two phases. The three events were defined as follows: (A and D) backswing end, corresponding to the minimum angular velocity of racket rotation in the horizontal plane; (B and E) ball contact, corresponding to the maximum resultant velocity of the racket; and (C and F) forward swing end, corresponding to the minimum angular velocity of racket rotation in the horizontal plane, as illustrated in Figure 3. To analyze the trajectory similarity between the backhand flick and the FRF, the entire racket movement process was divided into two phases: from the end of backswing to ball contact (A-B and D-E), and from ball contact to the forward swing end (B-C and E-F). Each trajectory was time-normalized to 101 equally spaced data points, corresponding to 0%–100% of the complete racket movement cycle.

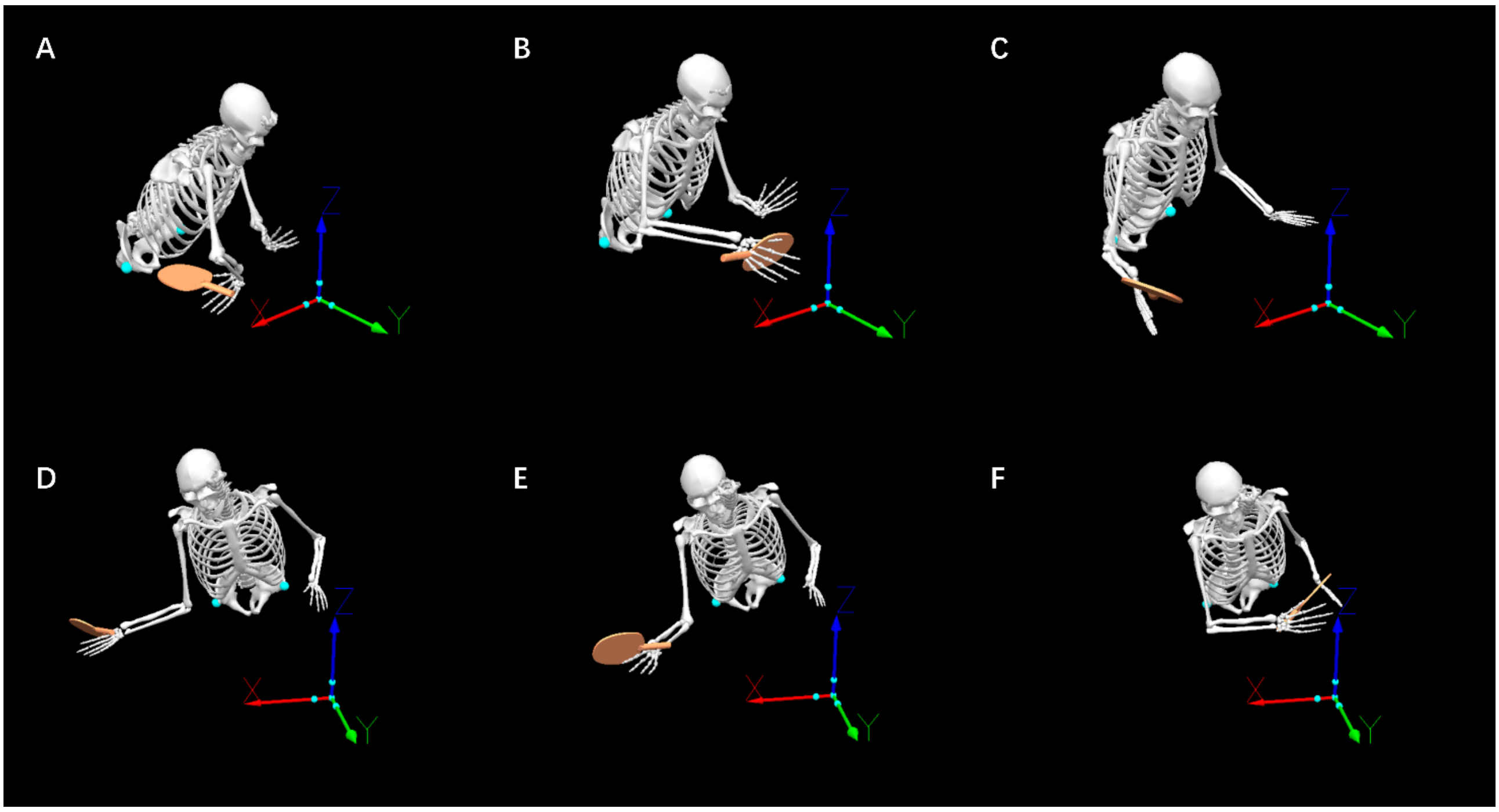

Figure 3 Definition of movement phases in the backhand flick and forehand reverse flick

**Data Processing**

Kinematic variables were processed using Visual 3D software (Visual 3D 6.0 Professional, C-Motion, Canada) with a Butterworth low-pass filter at a cutoff frequency of 15 Hz[14]. The laboratory coordinate system was defined as follows: the positive X-axis directed from the left greater trochanter center to the right greater trochanter center; the positive Y-axis perpendicular to the X-axis, pointing anteriorly from the midpoint of the two greater trochanters; and the positive Z-axis perpendicular to the X–Y plane, directed vertically upward. Joint motions followed the right-hand rule, and joint angles were calculated using the Cardan sequence recommended by the International Society of Biomechanics. To avoid gimbal lock in the wrist joint angle calculation, the Cardan sequence was set as X–Z–Y.

To eliminate positional and orientation differences between the two techniques, the Kabsch algorithm was used to optimally superimpose the two three-dimensional trajectories[15,16]. Only translation and rotation were permitted, and the algorithm solved for the rotation matrix and translation vector that minimized the root mean square deviation (RMSD), with a window length of 101 data points. The alignment procedure was defined as follows:

(1) For paired points P and Q of the two techniques, the covariance matrix was computed after centering the trajectories.

$$H = P^T Q$$

(2) Singular value decomposition (SVD) of the covariance matrix $H$:

$$H = USV^T$$

(3) The optimal rotation matrix is:

$$R = VU^T$$

If $\det(R) < 0$, the sign of the last column of $V$ is adjusted.

(4) The optimal translation vector is:

$$\mathrm{t} = \bar{Q} - R\bar{P}$$

(5) The aligned point set is:

$$P' = PR + \mathbf{t}$$

(6) The minimum root mean square deviation (RMSD) is calculated as:

$$\mathrm{RMSD} = \sqrt{\frac{1}{N}\sum_{i=1}^{N} \| P_i' - Q_i \|^2}$$

$P$, $Q$: two sets of paired three-dimensional points ($N \times 3$ matrices); $H$: covariance matrix after centering; $U, S, V$: matrices obtained from singular value decomposition; $R$: optimal rotation matrix; $\det(R)$: determinant of the rotation matrix; t: optimal translation vector; $P_i', Q_i$: coordinates of the $i$-th aligned points; $N$: number of points; $\|\|$: Euclidean norm; RMSD: root mean square deviation after alignment.

After completing Kabsch alignment (translation and rotation) to eliminate positional and directional differences, Procrustes analysis was further applied to the racket center of mass (COM) trajectory with isotropic scaling in order to compare the trajectory shape differences between the two techniques [17].

(1) Normalization of scale

$$\| X \|_F = \sqrt{\sum_{i=1}^{N}\sum_{j=1}^{3} x_{ij}^2}$$

$$P_s = \frac{P_c}{\|P_c\|_F},\ Q_s = \frac{Q_c}{\|Q_c\|_F}$$

Where $\| X \|_F$ denotes the Frobenius norm, defined as the square root of the sum of squared elements. $P_s$ and $Q_s$are the two trajectory matrices normalized to the same scale, $P_c$ and $Q_c$ are the centered trajectory coordinate matrices (centering was already performed in the kabsch alignment).

(2) Computation of shape disparity

$$\delta = 2 - 2 \cdot trace(\sum)$$

Where $\delta$ represents the Procrustes disparity, which is dimensionless, with $\delta$ = 0 indicating identical shapes. $trace(\sum)$ denotes the sum of singular values obtained from the singular value decomposition (SVD) in the Kabsch alignment.

### Statistical Analysis

The similarity of racket movement trajectories was analyzed using the Kabsch alignment algorithm implemented in Python 3.13. After alignment, Pearson correlation coefficients were calculated for the X, Y, and Z axes as well as for the overall trajectory. In addition, Procrustes disparity was applied to evaluate trajectory shape differences after isotropic scaling.

Discrete variables including racket vector velocity, resultant racket velocity, racket vector acceleration, and resultant racket acceleration were statistically analyzed using R 4.5.0. Given the small sample size ($n = 4$) and potential violation of normality assumptions, the nonparametric paired-sample Wilcoxon signed-rank test was employed to compare joint kinematic variables between the FRF and the backhand flick. To ensure accuracy of inference under small-sample conditions, exact p-values were reported without continuity correction. Effect sizes were expressed as rank-biserial correlations ($r$), and multiple comparisons were corrected using the Benjamini–Hochberg procedure. Data for each variable were expressed as median (Q1, Q3). For both techniques, descriptive and inferential statistics included quartiles, 95% confidence intervals, p-values, and effect sizes ($r$). The significance level for all analyses was set at $p < 0.05$.

## Results

### FRF success rate

Following seven consecutive days of training, the landing success rate of the FRF demonstrated a progressive improvement, as shown in Figure 4. On Day 1, the mean success rate was approximately 48%. Performance increased steadily across the subsequent four days, reaching approximately 75% by Day 5. On the final training day, the success rate peaked at approximately 86%.

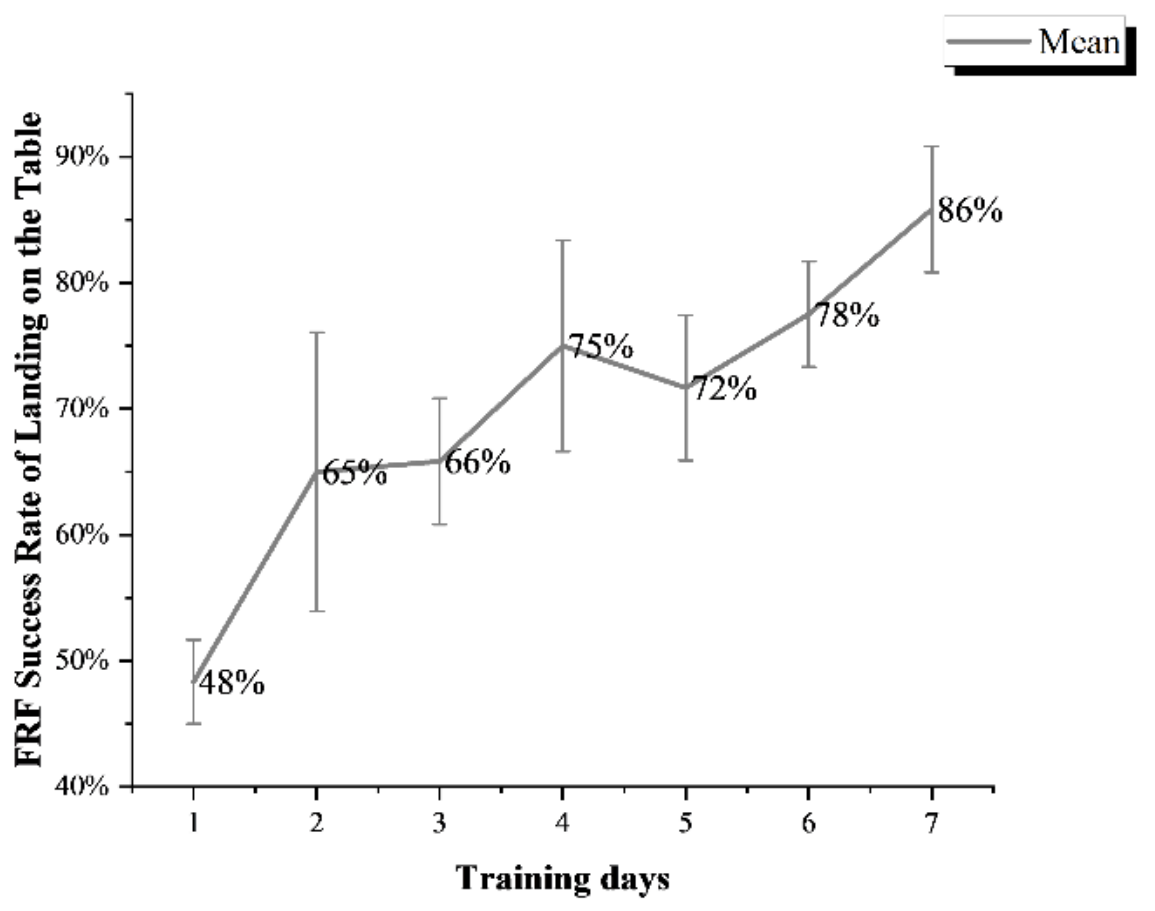

Figure 4 Landing success rate of the forehand reverse flick across seven consecutive training days

After normalization and alignment of the racket COM trajectories using the Kabsch algorithm, translation and rotation adjustments were applied. The alignment results for each participant are summarized in Table 2, and the FRF and BF three-dimensional trajectories before and after alignment are illustrated in Figure 5. Root mean square deviation (RMSD) values were notably reduced after alignment, indicating improved overlap between the FRF and BF trajectories. For the two techniques, Pearson correlation coefficients of the aligned trajectories demonstrated the highest correlation in the X-axis direction ($r = 1.00$), followed by the Y-axis ($r = 0.99$), whereas the Z-axis exhibited a negative correlation ($r = -0.04$). The overall three-dimensional correlation coefficient across axes was $r = 0.65$. Procrustes analysis, which included translation, rotation, and isotropic scaling, revealed an overall trajectory shape disparity of $2.49 \times 10^{-3}$.

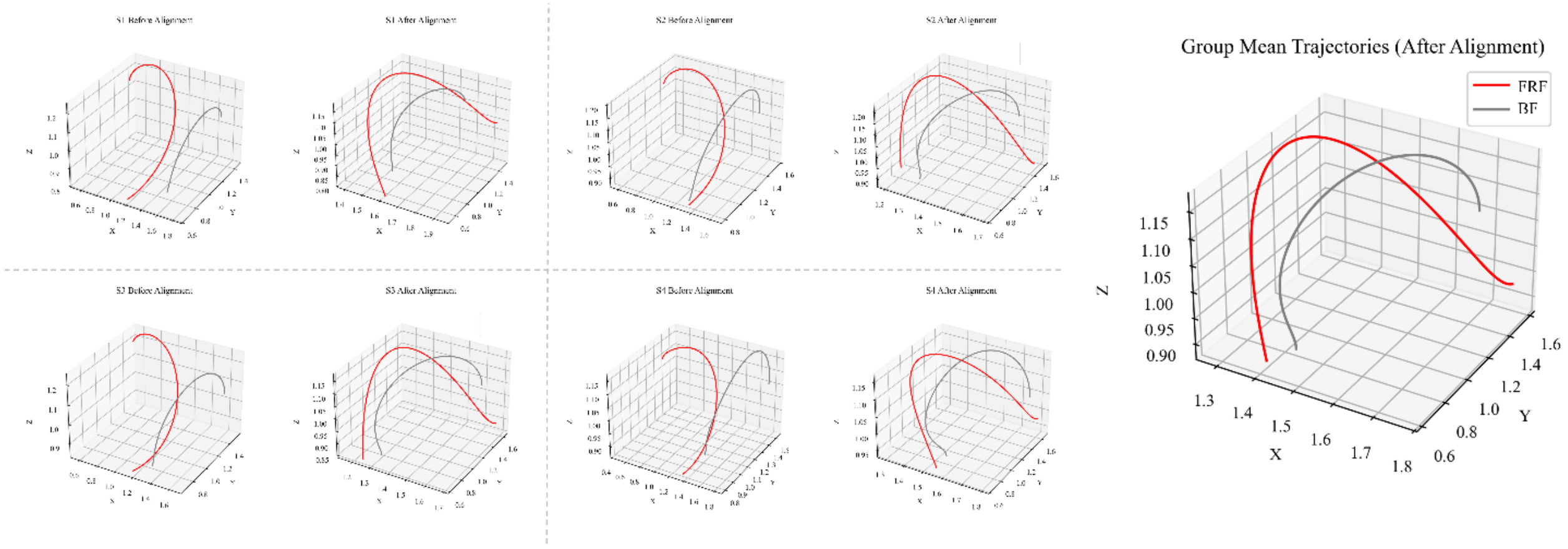

Figure 5 Three-dimensional trajectories of the racket COM before and after alignment

**Racket kinematics of the FRF and BF**

Table 2 Root mean square deviation, Pearson correlation coefficients, and Procrustes disparity of racket COM trajectories between the two techniques

| Racket COM Trajectory | RMSD before[m] | RMSD after[m] | Pearson X | Pearson Y | Pearson Z | Pearson mean | Procrustes disparity($\times 10^{-3}$) |
|---|---|---|---|---|---|---|---|
| *S*1 | 0.75 | 0.22 | 0.99 | 0.99 | 0.56 | 0.85 | 6.96 |
| *S*2 | 0.58 | 0.26 | 1.00 | 0.98 | -0.40 | 0.53 | 1.76 |
| *S*3 | 0.54 | 0.21 | 0.99 | 0.98 | 0.09 | 0.69 | 3.89 |
| *S*4 | 0.78 | 0.19 | 0.99 | 1.00 | -0.20 | 0.60 | 27.44 |
| Mean | 0.65 | 0.21 | 1.00 | 0.99 | -0.04 | 0.65 | 2.49 |

Note: Values of Procrustes disparity are scaled by $10^{-3}$ (i.e., multiply the tabulated value by 0.001 to obtain the raw value). Smaller values indicate greater trajectory similarity. *S* denotes subject, representing the mean value for each individual participant.

The racket COM velocities and accelerations relative to the laboratory coordinate system are shown in Table 3. Paired-sample Wilcoxon signed-rank tests revealed no significant differences between the two techniques for any velocity or acceleration variable ($p > 0.05$). Moderate effect size was observed in upward velocity ($r = 0.365$), while other variables demonstrated large effect sizes ($r \geqslant 0.73$). Regarding racket velocity, the FRF showed slightly higher peak values in the leftward (9.09 m/s), forward (9.31 m/s), and resultant directions (11.02 m/s) compared with the BF, but slightly lower upward velocity (4.75 m/s). For racket acceleration, the FRF demonstrated lower peak values across all three axes as well as in resultant acceleration, reaching 80.33% of the BF value.

Table 3 Peak racket COM velocities and accelerations

| Variables | Direction | Backhand Flick | Forehand Reverse Flick | 95%CI | *p* | *r* |
|---|---|---|---|---|---|---|
| Velocity | Left/Right [m/s] | 8.28(6.90,8.85) | 9.09(8.76,10.94) | [-2.57, -0.11] | 0.130 | 0.913 |
| | Forward [m/s] | 9.28(8.52,9.90) | 9.31(8.76,10.36) | [-0.53, 0.21] | 0.250 | 0.730 |
| | Upward [m/s] | 5.29(4.71,5.40) | 4.75(4.18,5.67) | [-0.67, 1.09] | 0.630 | 0.365 |
| | Resultant Velocity[m/s] | 10.50(9.56,10.85) | 11.02(10.58,12.36) | [-1.83, -0.49] | 0.130 | 0.913 |
| Acceleration | Left/Right [m/s$^2$] | 291.11(242.36,348.92) | 190.27(179.43,242.55) | [41.17, 107.48] | 0.125 | 0.913 |
| | Forward [m/s$^2$] | 155.97(152.05,189.64) | 114.12(106.88,163.32) | [21.89, 53.03] | 0.125 | 0.913 |
| | Upward [m/s$^2$] | 118.20(90.33,159.99) | 83.39(73.33,105.15) | [10.34, 80.11] | 0.125 | 0.913 |
| | Resultant Acceleration [m/s$^2$] | 330.60(292.51,382.91) | 265.57(247.16,317.58) | [39.65, 67.60] | 0.125 | 0.913 |

Note: Because the X axis movement directions of the two techniques were opposite, velocities and accelerations were defined as rightward positive for the backhand flick and leftward positive for the forehand reverse flick.

The durations of the two movement phases are presented in Table 4. Paired-sample Wilcoxon signed-rank tests revealed no significant differences between the two techniques in either phase ($p > 0.05$). However, effect sizes were large (r = 0.913). In both the backswing end to ball contact phase and the ball contact to forward swing end phase, the BF demonstrated shorter durations than the FRF, with a difference of approximately 0.03 s.

Table 4 Duration of two movement phases during racket swing (s)

| Movement Phases | Backhand Flick | Forehand Reverse Flick | 95%CI | *p* | *r* |
|---|---|---|---|---|---|
| Backswing end – Ball contact | 0.06(0.06,0.06) | 0.09(0.07,0.10) | [-0.04, -0.01] | 0.130 | 0.913 |
| Ball contact – Forward swing end | 0.09(0.07,0.09) | 0.12(0.11,0.14) | [-0.06, -0.02] | 0.130 | 0.913 |

**Ball velocity and spin rate of the FRF and BF**

The ball velocities and spin rates of the BF and FRF for each participant are illustrated in Figure 6, with overall values and Wilcoxon signed-rank test results summarized in Figure 7. No significant differences were found between the two techniques for either ball velocity or spin rate ($p > 0.05$). However, both variables demonstrated large effect sizes ($r > 0.5$). On average, the FRF achieved 96% of the BF velocity and 64% of the BF spin rate. At the individual level, the FRF outcomes were strongly correlated with the BF results, particularly for spin rate. For example, participant *S*4 achieved a maximum FRF spin rate of 102 r/s, with an individual mean of 93 r/s.

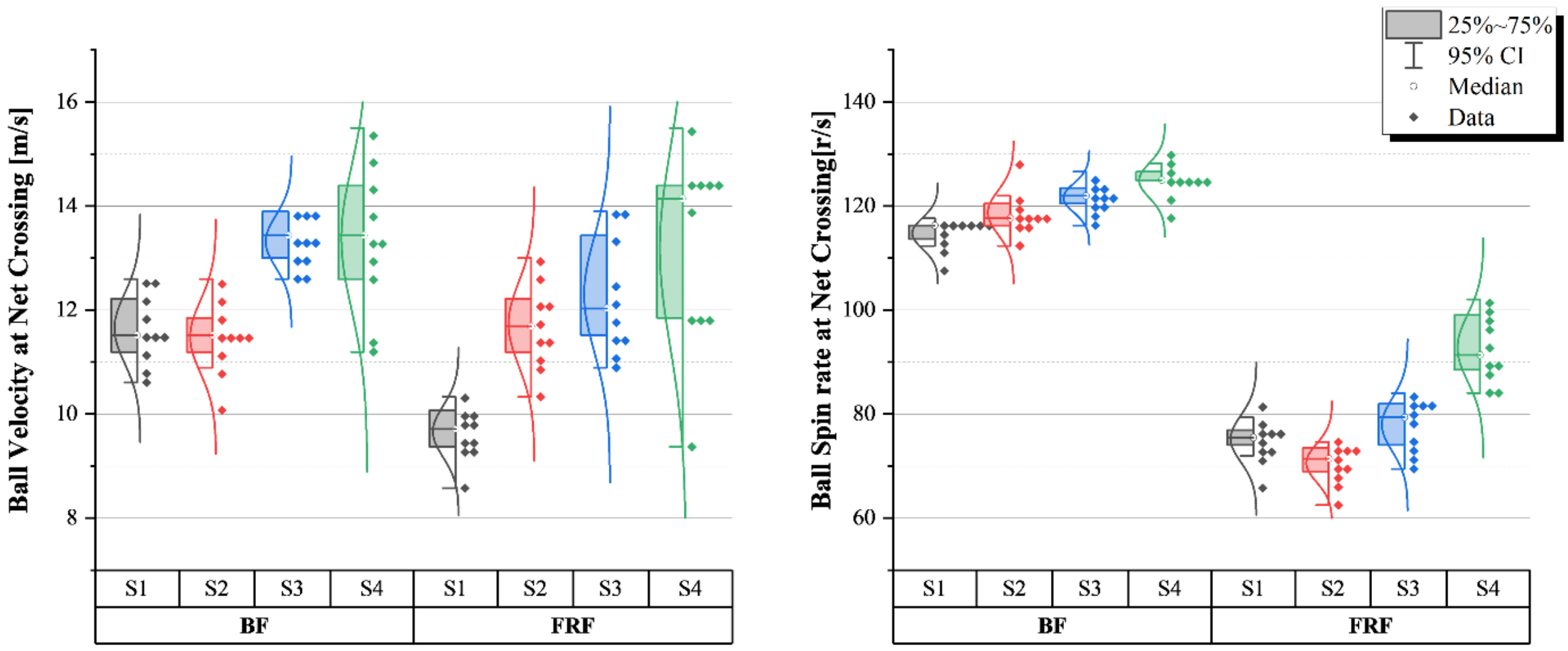

Figure 6 Ball velocity and spin rate of the BF and FRF for each participant

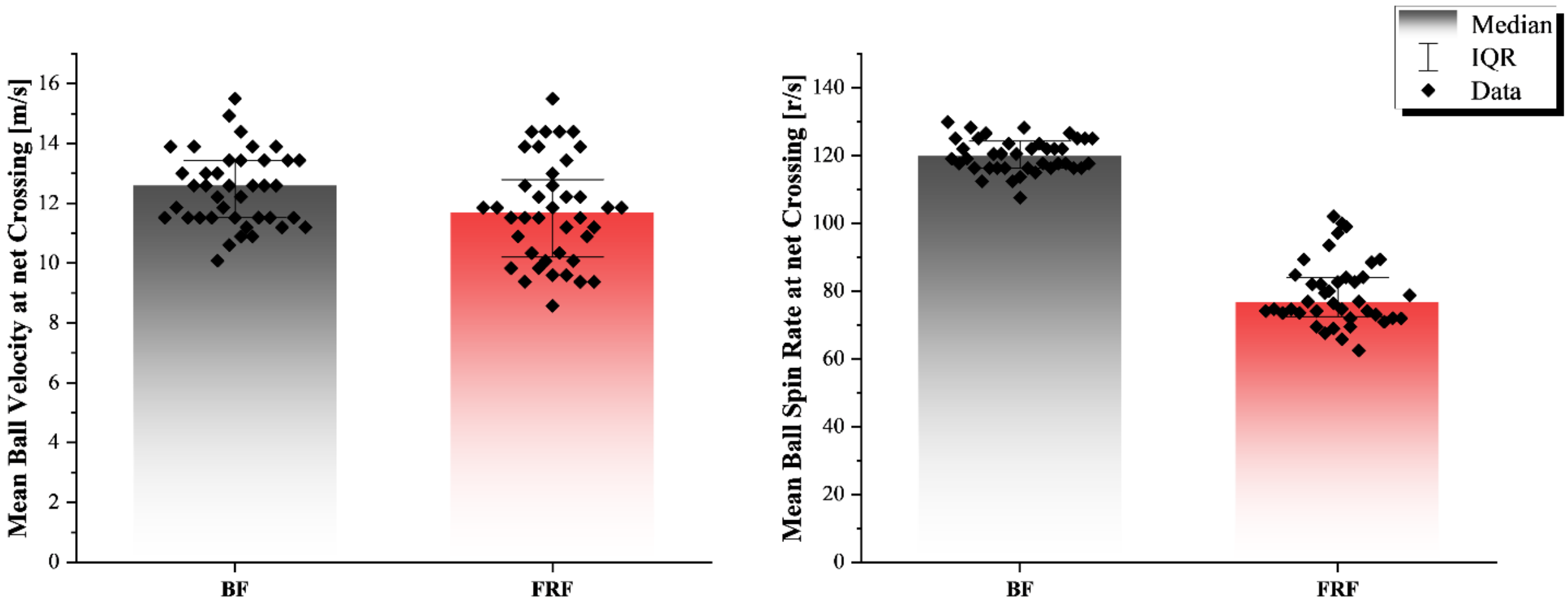

Figure 7 Mean ball velocity and spin rate of the BF and FRF

## Discussion

The purpose of this pilot study was to provide an initial verification of the feasibility of the FRF in table tennis and, for the first time, to investigate the kinematic characteristics and ball performance of elite athletes while learning and applying this technique. After seven days of training, the success rate of the FRF steadily increased, reaching approximately 86%, which supports Hypothesis 1 and indicates rapid learning potential and evidence of skill transfer. In terms of racket trajectory and velocity characteristics, the FRF exhibited a high degree of similarity with the BF, largely consistent with Hypothesis 2. Furthermore, the FRF achieved 96% of the ball velocity and 64% of the ball spin rate of the BF, which is in agreement with Hypothesis 3.

As a variant of the BF, the FRF technique demonstrated high similarity to the BF in terms of racket COM trajectory, racket velocity, and stroke timing. This overlap in movement goals and environmental information sources, according to the theories of affordances and specifying information[18–20], together with the view that elite athletes are more capable of achieving skill transfer[21], suggests that the FRF can be considered a case of near transfer from the BF. In this study, the success rate of the FRF steadily increased after seven days of practice, indicating both general motor learning processes and positive transfer from the BF. During training, athletes may have refined their stroke execution and body control strategies through external feedback in the form of knowledge of results, specifically provided by success rate testing, which facilitated the acquisition of the new technique[22].

Skill transfer was reflected not only in the improvement of the success rate but also in the similarity of kinematic characteristics. In the analysis of racket COM trajectories, the RMSD after Kabsch alignment was 0.21, and the disparity after Procrustes scaling was low, indicating a high degree of consistency in the overall movement patterns of the two techniques. Regarding racket velocity, both component velocities and resultant velocity of the FRF were comparable to those of the BF, suggesting that athletes were able to learn and adapt the new technique within a short training period to approach the velocity levels of the established technique. However, for acceleration variables, the FRF showed lower values than the BF, and longer durations were observed in both the backswing end–ball contact phase and the ball contact–forward swing end phase. This may reflect instability in determining stroke timing and contact point when executing the new technique, requiring extended movement duration to ensure successful performance[23]. Nevertheless, the ability of the FRF to approach the velocity magnitude of the BF indicates considerable plasticity, with the potential for further optimization of movement coordination and temporal control as training progresses.

Innovation in sport techniques requires not only kinematic validation of movement feasibility but, more importantly, an evaluation of their contribution to actual competitive performance[9,24,25]. In table tennis, technical execution ultimately acts on the ball, making ball velocity and spin rate critical indicators for assessing the competitive potential of new techniques[26,27]. For example, in the study of the forehand flip-loop (a newly developed with penhold grip technique inside the table), the maximum spin rate reached 85.4 r/s, which was higher than that of the traditional forehand flip (45.7 r/s) but slightly lower than the backhand flick (94.4 r/s). In terms of velocity, the forehand flip produced the highest values (10.91 m/s), compared with 10.25 m/s for the forehand flip-loop and 10.51 m/s for the backhand flick[12]. In the present study, the FRF produced ball velocities (12.02 m/s) comparable to the BF, indicating that this technique can achieve a similar offensive effect. For spin, the FRF reached 66% of the BF level (approximately 76.61 r/s) within a short training period, exceeding the traditional forehand flip (47.8 r/s)[6]. Notably, substantial inter-individual variability was observed: participant *S*4 demonstrated a mean FRF spin rate of 93 r/s, but with greater performance fluctuations[23,28]. This suggests that learning outcomes may differ depending on playing style and competitive level[29,30].

From a competitive perspective, the serve is one of the most tactical components in table tennis, with maximum spin rates reported at approximately 74.3 r/s[31]. The spin levels of the FRF have already approached and, in some cases, as illustrated by participant S4, who reached a maximum of 102 r/s, suggesting that the technique has potential in service return by counteracting or even surpassing the spin of the serve of opponent, thereby stabilizing the return and creating initiative for subsequent strokes. In summary, the FRF achieves sufficient ball velocity to sustain offensive threat while also ensuring the spin stability needed for reliable performance, thereby demonstrating considerable practical value in table tennis competition[27].

**Limitations and future directions**

This study has some limitations. First, although all participants were elite athletes, the sample size in this study was small ($n$ = 4). Therefore, the findings should be interpreted as preliminary and exploratory, serving primarily to verify the feasibility of the FRF technique. Second, the training period was limited to only seven days, which primarily reflects short-term learning effects and is insufficient to reveal the long-term stability and developmental potential of the technique. Future studies should expand the sample size and extend the training duration, while incorporating physiological measures and performance-related tactical indicators, to provide a more comprehensive evaluation of the competitive value of the technique.

## Conclusion

This study evaluated the performance characteristics and application potential of the forehand reverse flick (FRF), a forehand short return technique in table tennis. The findings indicate that elite athletes were able to acquire the FRF within a short training period, as reflected by a steady improvement in success rates. This rapid acquisition may be partly explained by the structural similarity between the FRF and backhand flick (BF), suggesting a potential role of skill transfer. The FRF exhibited racket center-of-mass trajectories, shape characteristics, racket velocities, and ball velocities that were comparable to those of the BF, although differences were observed in acceleration magnitude and phase duration. Regarding spin performance, the FRF generated lower spin rates than the BF. Overall, these findings suggest that the FRF is a technically attainable and practically promising return technique, with potential for further refinement and application in competitive table tennis.

**Acknowledgements**

Thanks to the Shanghai University of Sport and Table Tennis Intelligent R&D Center for providing excellent experimental conditions.

**Ethics Approval Disclosure**

The study protocol was approved by the Ethics Committee of Shanghai University of Sport (Approval No.102772025RT115).

**Consent for publication**

Prior to participation, participants were informed of the intent to publish the results of the study and provided written consent for publication.

**Author contributions**

P.J.: Conceptualization, Data curation, Methodology, Visualization, Writing – original draft. J.R.: Data curation, Investigation, Writing – review & editing. C.Y.: Data curation, Methodology. Q.K.: Conceptualization, Writing – review & editing. Q.Z.: Conceptualization, Resources. N.G.: Conceptualization, Writing – review & editing. B.C.: Conceptualization, Writing – review & editing. Z.F.: Conceptualization, Writing – review & editing.

**Funding**

This work was conducted without external funding.

**Declaration of conflicting interests**

The authors declared no potential conflicts of interest with respect to the research, authorship, and/or publication of this article.

**Data availability**

The datasets generated during and/or analyzed during the current study are available from the corresponding author on reasonable request.